\documentclass[lettersize,journal]{IEEEtran}
\usepackage{amsmath,amsfonts}
\usepackage{algorithmic}
\usepackage{algorithm}
\usepackage{array}
\usepackage[caption=false,font=normalsize,labelfont=sf,textfont=sf]{subfig}
\usepackage{textcomp}
\usepackage{stfloats}
\usepackage[hyphens]{url}
\usepackage{verbatim}
\usepackage{orcidlink}
\usepackage{graphicx}
\usepackage{cite}
\hypersetup{breaklinks=true}

\newcommand{\emissionsOptimisticScenario}{150.5\ }
\newcommand{\emissionsPessimisticScenario}{994.9\ }
\newcommand{\emissionsLikelyScenario}{456.9\ }
\newcommand{\waterOptimisticScenario}{1,989.6\ }
\newcommand{\waterPessimisticScenario}{37,664.5\ }
\newcommand{\waterLikelyScenario}{5,738.2\ }
\newcommand{\urbanTreesLikelyScenario}{7,615\ }
\newcommand{\averageAmericanLikelyScenario}{5,053\ }
\newcommand{\smartphonesLikelyScenario}{70,635,452\ }
\newcommand{\showersLikelyScenario}{94,738\ }
\newcommand{\workflowRuns}{2,226,729\ }
\newcommand{\repositories}{18,683\ }
\newcommand{\frangos}{3,050,167\ }
\newcommand{\glasses}{22,953,162\ }

\usepackage{paralist}
\usepackage{listings}
\usepackage[breakable]{tcolorbox}
\usepackage{lipsum}
\usepackage{MnSymbol}
\usepackage{multicol}
\usepackage{enumitem}
\usepackage{cuted}
\usepackage[group-separator={,}]{siunitx}
\newcommand{\TODO}[1]{\textcolor{red}{#1}\GenericWarning{}{LaTeX Warning: TODO: #1}}\newcommand\todo\TODO

\usepackage{titlesec}
\titleformat{\subparagraph}[runin]{\normalfont\normalsize\bfseries}{\thesubparagraph}{1em}{}
\titlespacing*{\subsubsection}{\parindent}{2ex plus 1ex minus .2ex}{0.5ex plus .2ex}
\titlespacing*{\paragraph}{2\parindent}{0.5ex plus 1ex minus .2ex}{0.5ex plus .2ex}
\titlespacing*{\subparagraph}{3\parindent}{0.5ex plus 1ex minus .2ex}{0.5ex plus .2ex}

\newcommand\YAMLcolonstyle{\color{red}\mdseries}
\newcommand\YAMLkeystyle{\color{black}\bfseries}
\newcommand\YAMLvaluestyle{\color{blue}\mdseries}

\makeatletter

\newcommand\language@yaml{yaml}

\expandafter\expandafter\expandafter\lstdefinelanguage
\expandafter{\language@yaml}
{
  keywords={true,false,null,y,n},
  keywordstyle=\color{darkgray}\bfseries,
  basicstyle=\YAMLkeystyle,
  sensitive=false,
  comment=[l]{\#},
  morecomment=[s]{/*}{*/},
  commentstyle=\color{purple}\ttfamily,
  stringstyle=\YAMLvaluestyle\ttfamily,
  moredelim=[l][\color{orange}]{\&},
  moredelim=[l][\color{magenta}]{*},
  moredelim=**[il][\YAMLcolonstyle{:}\YAMLvaluestyle]{:},
  morestring=[b]',
  morestring=[b]",
  literate =    {---}{{\ProcessThreeDashes}}3
                {>}{{\textcolor{red}\textgreater}}1
                {|}{{\textcolor{red}\textbar}}1
                {\ -\ }{{\mdseries\ -\ }}3,
}

\lst@AddToHook{EveryLine}{\ifx\lst@language\language@yaml\YAMLkeystyle\fi}
\makeatother

\newcommand\ProcessThreeDashes{\llap{\color{cyan}\mdseries-{-}-}}

\begin{document}

\title{Environmental Impact of CI/CD Pipelines}

\author{
    Nuno Saavedra\orcidlink{0000-0003-4148-5991}$^\ast$,
    Alexandra Mendes\orcidlink{0000-0001-8060-5920},
    João F. Ferreira\orcidlink{0000-0002-6612-9013}
    \IEEEcompsocitemizethanks{
            \IEEEcompsocthanksitem N. Saavedra and J. F. Ferreira are with INESC-ID and IST, University of Lisbon, Portugal.\protect\\
		E-mail: nuno.saavedra@tecnico.ulisboa.pt, joao@joaoff.com
            \IEEEcompsocthanksitem A. Mendes is with INESC TEC and Faculty of Engineering, University of Porto, Portugal. \protect\\
        E-mail: alexandra@archimendes.com

        \protect}
    \thanks{$^\ast$Corresponding author.}
}

\markboth{Journal of \LaTeX\ Class Files,~Vol.~14, No.~8, August~2021}
{Shell \MakeLowercase{\textit{et al.}}: A Sample Article Using IEEEtran.cls for IEEE Journals}

\maketitle

\begin{abstract}
Continuous Integration and Continuous Delivery (CI/CD) pipelines are widely used in software development, yet their environmental impact,
particularly carbon and water footprints (CWF), remains largely unknown to developers, as CI service providers typically do not disclose such information.
With the growing environmental impact of cloud computing, understanding the CWF of CI/CD services has become increasingly important.

This work investigates the CWF of using GitHub Actions, focusing on open-source repositories where usage is free and unlimited for standard runners.
We build upon a methodology from the Cloud Carbon Footprint framework and we use the largest dataset of workflow runs reported in the literature to date,
comprising over 2.2 million workflow runs from more than 18,000 repositories.

Our analysis reveals that the GitHub Actions ecosystem results in a substantial CWF. Our estimates for the carbon footprint in 2024 range from \emissionsOptimisticScenario MTCO2e in the
most optimistic scenario to \emissionsPessimisticScenario MTCO2e in the most pessimistic scenario, while the water footprint ranges from \waterOptimisticScenario to \waterPessimisticScenario kiloliters. The most likely scenario
estimates are \emissionsLikelyScenario MTCO2e for carbon footprint and \waterLikelyScenario kiloliters for water footprint. To provide perspective, the carbon footprint in the most likely scenario is
equivalent to the carbon captured by \urbanTreesLikelyScenario urban trees in a year, and the water footprint is comparable to the water consumed by an average American family over \averageAmericanLikelyScenario years.

We explore strategies to mitigate this impact, primarily by reducing wasted computational resources. Key recommendations include deploying runners in regions whose energy production has a low environmental impact such as France and the United Kingdom, implementing stricter deactivation policies for scheduled runs and aligning their execution with periods when the regional energy mix is more environmentally favorable, and reducing the size of repositories.

This study provides crucial insights into the environmental impact of CI/CD runs and offers a foundation for future sustainability efforts in this domain.
\end{abstract}

\begin{figure}
    \centering
    \includegraphics[width=\linewidth]{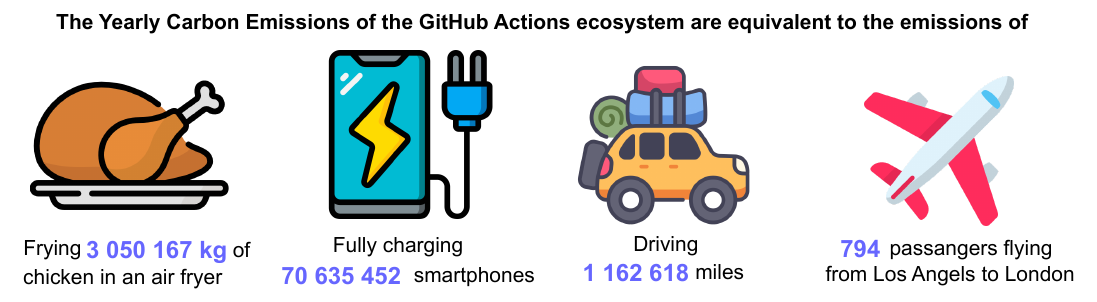}
    \caption{Comparison between the yearly carbon emissions of the GitHub Actions ecosystem and the emissions of quotidian activities~\cite{EPA-2024,baumeister2017each,Rousseau_2022}.}
    \label{fig:emissions_comparison}
\end{figure}

\begin{figure}
    \centering
    \includegraphics[width=0.8\linewidth]{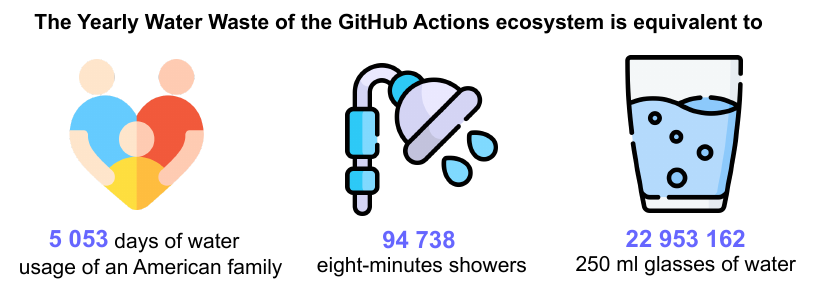}
    \caption{Comparison between the yearly water waste of the GitHub Actions ecosystem and the water usage of quotidian activities~\cite{EPA-2025a, EPA-2025b}.}
    \label{fig:water_comparison}
\end{figure}


\section{Introduction}
\label{sec:introduction}

Continuous Integration and Continuous Delivery (CI/CD) are software development practices that enable fast iteration of software versions. CI allows developers to receive fast feedback, allowing them to quickly know if their changes integrate with existing code~\cite{gallaba2022lessons}. CD ensures that software can be reliably released at any time~\cite{chen2015continuous}, usually in a fully automated way. CI/CD pipelines are a consistent process comprising the steps required to build, test, review, integrate, and deliver software artifacts. CI/CD pipelines can be automated by using services such as GitHub Actions~\cite{githubactions_2025}, TravisCI~\cite{travisci_2025}, and CircleCI~\cite{circleci_2025}. These automated pipelines are started manually or triggered by development events, such as the creation of a pull request or the push of a commit to the main branch.

Despite the benefits of automated CI/CD pipelines, their usage raises the question: what are the costs associated with automated CI/CD pipelines? The first cost that typically comes to mind is the financial cost. Previous studies have explored the financial costs of CI/CD pipelines and how to optimize resource usage to reduce these costs~\cite{bouzenia2024resource}.
CI/CD service providers, such as GitHub Actions, bill their users according to the execution time of their CI/CD pipelines.
However, CI/CD service providers may also support a free tier option where organizations are given a set amount of free minutes per month to execute their pipelines.
In particular, GitHub Actions allows free and unlimited usage of their services for open-source repositories\footnote{Limited to standard GitHub-hosted runners.}.

In this case, a critical but often overlooked cost is energy consumption. Automated CI/CD pipelines, as any other computer program, require hardware to run, which consequently requires energy. Therefore, when developers trigger a CI/CD pipeline, they start a computation that consumes energy. However, developers are usually not aware of the energy their pipelines consume, since CI/CD service providers do not provide information about energy consumption. Moreover, pipelines can be scheduled or automatically triggered by development events, which may make developers unaware of the execution of their pipelines.

As long as energy production remains dependent on non-carbon-neutral sources, the execution of CI/CD pipelines will inevitably have an environmental impact. Two key measures of this impact are the carbon footprint and the water footprint. Wiedmann et al. define carbon footprint as ``a measure of the exclusive total amount of carbon dioxide emissions that are directly and indirectly caused by an activity or accumulate over the life stages of a product''~\cite{wiedmann2008definition}. Similarly, Hoekstra et al. define the water footprint of a product as ``the volume of freshwater used to produce the product, measured over the full supply chain''~\cite{hoekstra2012water}.

In this work, we explore the carbon and water footprints (CWF) of using CI/CD pipelines in software projects.

Following the definitions of Wiedmann et al. and Hoekstra et al., we consider not only the direct costs of executing CI/CD pipelines, such as the energy consumed during the pipelines execution, but also the indirect costs.

Indirect costs include the environmental costs associated with hardware manufacturing and, for example, the freshwater consumption required for cooling data centers~\cite{jiang2025waterwise,CloudCarbonFootprint-2024,Green_Software_Foundation_Software_Carbon_Intensity_2024}.

We use the ecosystem of open-source repositories using GitHub Actions as our case study. There are three reasons for our choice:
\begin{inparaenum}
    \item GitHub Actions is currently one of the most popular CI/CD service providers~\cite{JetBrains_2023};
    \item the data related to GitHub Actions pipeline runs for open-source repositories is publicly available;
    \item open-source projects are not billed when using GitHub Actions which might reduce the incentive to optimize the execution time of CI/CD pipelines.
\end{inparaenum}

After understanding what the environmental impact of the GitHub Actions pipelines is, we explore strategies to reduce it.
We focus on avoiding wasted computational resources and how CI/CD service providers can help reduce the environmental impact of their services.

We structure our study by addressing the following research questions.

\begin{tcolorbox}[colback=gray!20!white,colframe=gray!70!black,title=RQ1: What is the carbon and water footprints of the GitHub Actions ecosystem?,arc=10pt]
We used the year 2024 as a case study to evaluate the environmental impacts associated with the GitHub Actions ecosystem.
In 2024, our estimates for the carbon footprint of the GitHub Actions ecosystem range from \emissionsOptimisticScenario MTCO2e\footnote{MTCO2e stands for Metric Tons of Carbon Dioxide Equivalent.} in the most optimistic scenario to \emissionsPessimisticScenario MTCO2e in the most pessimistic scenario, while the water footprint ranges from \waterOptimisticScenario to \waterPessimisticScenario kiloliters. In the most likely scenario, the carbon footprint is estimated at \textbf{\emissionsLikelyScenario MTCO2e} and the water footprint at \textbf{\waterLikelyScenario kiloliters}, equivalent to the emissions of fully charging \smartphonesLikelyScenario smartphones and the water consumption of \showersLikelyScenario eight-minute showers. Figures~\ref{fig:emissions_comparison} and~\ref{fig:water_comparison} compare the CWF of the GitHub Actions ecosystem with other quotidian activities.
\end{tcolorbox}

\begin{tcolorbox}[colback=gray!20!white,colframe=gray!70!black,title=RQ2: What are effective strategies to reduce the environmental impact of the GitHub Actions ecosystem?,arc=10pt]
To reduce the carbon footprint of the GitHub Actions ecosystem, effective strategies include deploying runners in regions whose energy production has a low environmental impact such as France and the United Kingdom, implementing stricter deactivation policies for scheduled runs, aligning their execution with periods when the regional energy mix is more environmentally favorable, and optimizing repository cloning by reducing repository sizes. Another strategy could be to enhance transparency by displaying the carbon and water footprints of workflow runs to developers. Providing comparative metrics of the carbon and water footprints between users and repositories can further encourage sustainable practices.
\end{tcolorbox}

In summary, our contributions are as follows:
\begin{enumerate}
    \item a quantification of the CWF of the GitHub Actions ecosystem, providing critical insights into the environmental impact of CI/CD runs and serving as a foundation for future sustainability efforts;
    \item a dataset of \workflowRuns workflow runs from \repositories different public repositories actively using GitHub Actions in 2024, which can support further research and replication by the community. To the best of our knowledge, this is the largest dataset of workflow runs in the literature;
    \item effective strategies to reduce the CWF of the GitHub Actions ecosystem, which, if adopted, can lead to significant reductions in the environmental impact of GitHub Actions.
\end{enumerate}

\section{GitHub Actions}
A CI/CD pipeline is a sequence of automated processes designed to build, test, or deploy new versions of software efficiently and reliably. In this paper, we focus on GitHub Actions, one of the most popular CI/CD platforms~\cite{JetBrains_2023}. In GitHub Actions, developers write scripts that define each step executed by the pipeline. These scripts are called workflows. Figure~\ref{fig:workflow_example} shows a simplified version of the workflow defined to test the Flacoco fault localization tool.

\begin{figure}
\centering
\hfill
\begin{minipage}{0.455\textwidth}
\begin{lstlisting}[language=Yaml, frame=single, numbers=left, basicstyle=\scriptsize]
name: tests
on: [push, pull_request]
jobs:
  build:
    runs-on: ${{ matrix.os }}
    strategy:
      matrix:
        java-version: [11, 17]
        compiler-version: [12, 13, 14, 15, 17]
        os: [
          ubuntu-latest, macos-latest, windows-latest
        ]
    steps:
      - uses: actions/checkout@v4.2.2
      - name: Setup JDK${{ matrix.java-version }}
        uses: actions/setup-java@v4.6.0
        with:
          java-version: ${{ matrix.java-version }}
          distribution: 'temurin'
      - name: Install example projects
        run: ./.github/install_examples.sh
        env:
          SRC_VERSION: ${{ matrix.compiler-version }}
      - name: Build and run tests
        run: mvn --batch-mode clean test
        env:
          SRC_VERSION: ${{ matrix.compiler-version }}
      - name: Codecov
        uses: codecov/codecov-action@v5.1.2
\end{lstlisting}
\end{minipage}
\caption{Simplified version of the workflow defined to test Flacoco\protect\footnotemark, a fault localization tool for Java.}
    \label{fig:workflow_example}
\end{figure}

\footnotetext{\url{https://github.com/ASSERT-KTH/flacoco/blob/bc5d23d11b7afc24fc7a2fe1fd072f58e2322e9e/.github/workflows/tests.yml}}

Line 2 defines the triggers of the workflow. A trigger is the event that starts a workflow run. The workflow in Figure~\ref{fig:workflow_example} runs every time a push or pull request is made. Each workflow run can have multiple jobs (line 3). Each job runs in a runner environment specified by the \textit{runs-on} attribute (line 5). Developers can define multiple settings for the same job, generating a new job for each setting. Lines 7 to 12 define a matrix of all possible settings for the \textit{build} job. In our example, each setting has a different combination of Java version, compiler version, and runner environment. For each job, the developer must specify the steps to execute (lines 13 to 29). For each step, the developer can specify either an action to use (line 14) or a shell command to run (line 21). An action is an abstraction that encapsulates the execution of a complex and repetitive task. For instance, the action used on line 14 clones the repository on which the workflow is running.

\section{Carbon Footprint Estimation}
\begin{figure*}[ht]
    \centering
    \includegraphics[width=0.9\linewidth]{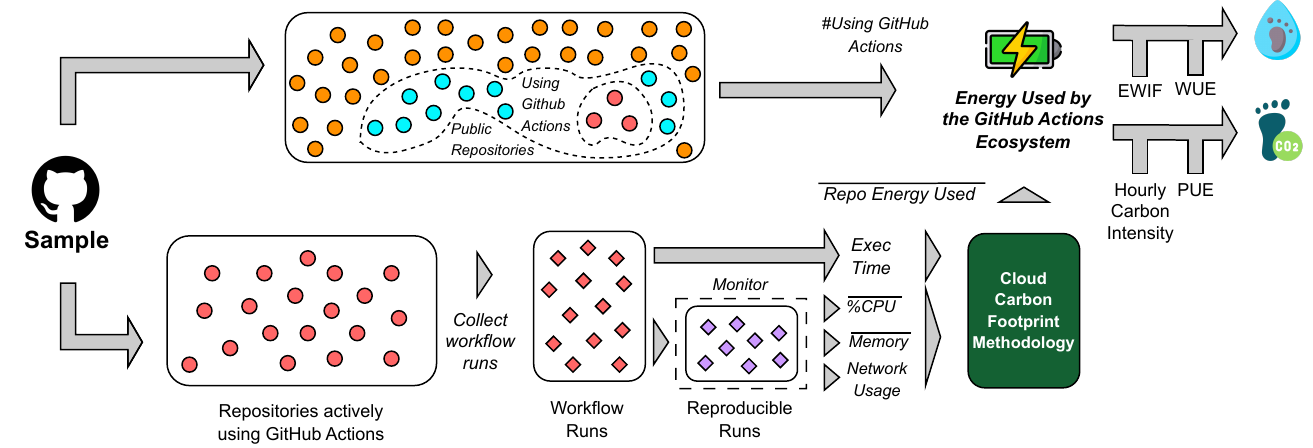}
    \caption{Overview of the methodology to calculate the carbon and water footprints of the GitHub Actions Ecosystem.}
    \label{fig:methodology}
\end{figure*}

To answer \textit{RQ1}, we calculate the estimated CWF of the entire GitHub Actions ecosystem for 2024. The answer to \textit{RQ2} comes from the analysis of the results and data collected for the first research question.

At the time of writing, GitHub reported that it has more than 420 million repositories~\cite{Github_2024}. Even if only a small portion of these repositories use GitHub Actions, the amount of data would be impractical for our study. For this reason, we rely on estimations calculated from a sample of all public repositories on GitHub. To calculate an estimate of the CWF of the entire GitHub Actions ecosystem for 2024, we need to estimate:

\begin{enumerate}
    \item the number of public repositories actively using GitHub Actions ($R_{GA}$);
    \item the average yearly carbon and water footprints of an active repository using GitHub Actions ($\overline{C_f}$ and $\overline{W_f}$).
\end{enumerate}

The methodology for each of these estimates is explained in Sections~\ref{sec:active-repos} and~\ref{sec:average-carbon}, respectively. We multiply $\overline{C_f}$ by $R_{GA}$ to obtain the yearly carbon footprint and $\overline{W_f}$ by $R_{GA}$ to obtain the yearly water footprint. An overview of our methodology is shown in Figure~\ref{fig:methodology}.

\subsection{Repositories actively using GitHub Actions}
\label{sec:active-repos}

To estimate the number of public and active repositories actively using GitHub Actions, we need to know the total number of repositories on GitHub ($R_{total}$). Kashyap analyzed the GitHub repository IDs and concluded that these IDs are incremental and shared between public and private repositories~\cite{Kashyap_2024}. Using this information, to get $R_{total}$ at any particular moment, we collect the last repository created up to that moment and extract its ID. As of the last day of 2024, $R_{total}$ reached 910,652,743.

We can multiply $R_{total}$ by the proportion of public and active repositories actively using GitHub Actions ($R_{GA_\%}$) to obtain $R_{GA}$. We consider a repository to be active if it is not archived and to be actively using GitHub Actions if the GitHub API\footnote{\url{https://docs.github.com/en/rest?api}} returns at least one workflow run for the year 2024. To obtain $R_{GA_\%}$, we must collect a random sample from the entire repository population.

To collect each repository of the sample, we randomly choose an ID between 0 and $R_{total}$. We call the GitHub API to get the repository with the chosen ID. If we cannot retrieve the repository, we know that the repository is private or has been deleted. Otherwise, the repository is public. If the repository is public, we check if it has been archived. If not, we count it as a public and active repository. Then, we check if the repository is actively using GitHub Actions. We keep collecting repositories until we get 20,001 repositories actively using GitHub Actions. To our knowledge, only one prior study includes a larger number of GitHub repositories, but it conducts a static analysis of workflow definitions and does not examine actual executions~\cite{decan2022use}. In contrast, studies like ours that analyze workflow executions have so far considered significantly smaller samples of just 10 and 952 repositories~\cite{bouzenia2024resource,classen2023carbon}.

Our final sample comprises 1,646,552 repositories, including 626,637 public and active repositories ($\approx$38.1\%), of which 20,001 actively use GitHub Actions ($\approx$1.2\%). Multiplying these proportions by $R_{total}$ yields an estimated total of 346,571,929 public and active repositories, with a margin of error of ±0.097\% at a 99\% confidence interval, and 11,061,883 repositories actively using GitHub Actions ($R_{GA}$), with a margin of error of ±0.022\% at a 99\% confidence interval. To calculate the error margin, we applied the standard formula for finite populations, considering $R_{total}$ as the population size, a sample size of 1,646,552, and the observed proportions of public and active repositories or repositories using GitHub Actions.

\subsection{Average Carbon and Water Footprints of a Repository}
\label{sec:average-carbon}

As GitHub does not provide data on neither the energy consumption nor the CWF of the GitHub Actions runs, we must estimate the average carbon and water footprints of an active repository using GitHub Actions ($\overline{C_f}$ and $\overline{W_f}$). We build upon the methodology of the Cloud Carbon Footprint application~\cite{CloudCarbonFootprint-2024} to calculate the energy used by the GitHub Actions ecosystem.
Cloud Carbon Footprint is an open-source project, sponsored by Thoughtworks Inc., that estimates the carbon emissions of using public cloud providers such as AWS, Azure, and GCP. It has been used to measure and reduce cloud carbon emissions of organizations such as Thoughtworks and OSP~\cite{CloudCarbonFootprintAdopters-2024}. Given that GitHub-hosted runners for GitHub Actions are deployed on Azure~\cite{GithubRunners-2024}, the Cloud Carbon Footprint methodology can be used.
Then, we use the estimated energy consumption from the Cloud Carbon Footprint methodology to calculate the carbon footprint, but instead of using the calculation from the original methodology, we modified this calculation to improve its accuracy. Additionally, we adapted the methodology to include water footprint estimation.

\subsubsection{Cloud Carbon Footprint Methodology}
\label{sec:cloud-carbon-footprint}

\noindent\indent In this section, we explain the components of the Cloud Carbon Footprint methodology that we use, how we apply them, and the modifications we made. Our explanation follows the original description of this methodology~\cite{CloudCarbonFootprint-2024}. The estimate of total carbon emissions is given by the following formula:

\begin{align}
    \textit{Carbon Footprint}
    &= \textit{Operational Emissions} \nonumber \\
    &\phantom{=} \quad +\ \textit{Embodied Emissions}
\label{eq:total_co2}
\end{align}

In our case, the operational emissions are related to the execution of the GitHub Actions pipelines, that is, the emissions caused by the production of the energy consumed for executing the GitHub Actions pipelines. The embodied emissions are the carbon emissions related to the manufacturing of the hardware required to execute the GitHub Actions pipelines. This formula is in accordance with the Software Carbon Intensity Specification of the Green Software Foundation~\cite{Green_Software_Foundation_Software_Carbon_Intensity_2024}.

\paragraph{Operational Emissions} The operational emissions are calculated by\footnote{Units of measurement are enclosed in square brackets (e.g., [kWh]) to provide additional context and improve clarity for the reader when interpreting formulas.}:

\begin{align}
    &\textit{Operational Emissions} = \notag \\
    &\qquad\begin{aligned}
        &\phantom{\times~}\textit{PUE} \\
        &\times~\textit{Grid Emission Factors}~[MTCO2e]\\ &\times~\textit{Cloud Energy Consumption}~[kWh]
    \end{aligned}
\label{eq:operation_emissions}
\end{align}

PUE corresponds to the \textit{Cloud provider Power Usage Effectiveness}.
The Cloud Carbon Footprint team fixes the PUE at $1.185$~\cite{CloudCarbonFootprint-2024}, which is the global value provided by the Microsoft Sustainability team. However, we use the region-specific values provided by the Microsoft Sustainability team: 1.17 for the Americas, 1.405 for Asia Pacific, and 1.185 for Europe, the Middle East and Africa~\cite{Noelle2022}.
The \textit{Grid Emission Factors} represent the average carbon emissions associated with the consumption of electricity from the grid and it depends on the region in which Azure instances are deployed. However, GitHub does not disclose the region where runners are deployed. In 2021, the GitHub support stated that runners were deployed exclusively in the United States~\cite{GithubActionsRegions-2021}, but this information is not official. Given the uncertainty about the regions used, we consider multiple regions around the world in our study. This not only will highlight the importance of deploying these services in carbon-aware regions, but also gives us an indication about the possible scenarios.
In the Cloud Carbon Footprint Methodology, the values for the \textit{Grid Emission Factors} are calculated by the Cloud Carbon Footprint team or by institutions such as the US Environmental Protection Agency (EPA) and are fixed over time according to the region.

However, grid emission factors vary over time, making it more realistic to use the grid emission factor at the time energy is consumed.
To account for this variability, our approach uses historical carbon intensity data with hourly granularity for a given region, obtained from Electricity Maps~\cite{ElectricityMaps2025}, as an estimate of grid emission factors. Carbon intensity quantifies the amount of carbon emissions produced per unit of electricity generated. Considering the average carbon intensity of the grid, we can use it as a proxy for the \textit{Grid Emission Factors}.
For each workflow run, we use the hourly carbon intensity of the specific hour in which the workflow run started.

The \textit{Cloud Energy Consumption} is calculated as follows:

\begin{align}
    &\textit{Cloud Energy Consumption} = \nonumber \\
    &\qquad\begin{aligned}
        & \textit{Compute}_\lightning\ +\ \textit{Storage}_\lightning\ +\ \textit{Memory}_\lightning\ +\ \textit{Network}_\lightning
    \end{aligned}
\label{eq:cloud_energy}
\end{align}

The methodology for both Equations~\ref{eq:operation_emissions} and \ref{eq:cloud_energy} was based on Etsy's Cloud Jewel approach~\cite{Etsy_2020}, with the addition of considerations for network and memory usage. The formulas to calculate each component of Equation~\ref{eq:cloud_energy} are stated below.

\allowdisplaybreaks
\begin{align}
\textit{Compute}_\lightning\ =\nonumber\\ \bigl(~vCPU_{min_\lightning}\ [kW]\ \times\ \#vCPUs\ ~~+\nonumber\\ (vCPU_{max_\lightning}\ [kW]\ -\ vCPU_{min_\lightning})\ \times\ \overline{vCPU_\textit{usage}}~\bigr) \nonumber\\\times\ \textit{Exec Time}~[h]
\label{eq:compute}\\\nonumber\\
\textit{Storage}_\lightning\ =\nonumber\\ \textit{Storage Coefficient}~\Bigr[\frac{kWh}{TB/h}\Bigr] \nonumber\\ \times\ \textit{Reserved Storage}~[TB] \times \textit{Exec Time}~[h]
\label{eq:storage}\\\nonumber\\
\textit{Memory}_\lightning\ =\nonumber\\
\textit{Memory Coefficient}~\Bigr[\frac{kWh}{GB/h}\Bigr] \nonumber\\
\times\ \overline{\textit{Memory}_\textit{usage}}\ [GB]\ \times\ \textit{Exec Time}~[h]
\label{eq:memory}\\\nonumber\\
\textit{Network}_\lightning =\nonumber\\
\textit{Network Coefficient}\ \Bigr[\frac{kWh}{GB}\Bigr]\ \times\ \textit{Network}_\textit{usage}\ [GB]
\label{eq:network}
\end{align}

\subparagraph{$\mathbf{Compute_\lightning}$.} Equation~\ref{eq:compute} describes how we calculate the energy spent on computing.
The metric $\overline{vCPU_\textit{usage}}$ represents the mean number of virtual CPUs used during a GitHub Actions run. The calculation of this metric is detailed in Section~\ref{sec:exec_metrics}.
Usually, when using the cloud, instead of physical CPUs, processes are executed on vCPUs. The energy consumed by a virtual CPU is usually lower than that of a physical CPU, since multiple virtual CPUs can run on a single physical CPU.
GitHub Actions reserves a virtual machine with 4 vCPUs for each job~\cite{GithubRunners-2024} ($\#vCPUs$). Each of these vCPUs will continuously consume a minimum amount of energy independently of the process being executed ($vCPU_{min_\lightning}$)~\cite{CloudCarbonFootprint-2024,Etsy_2020}. When the vCPU is being fully used, it consumes $vCPU_{max_\lightning}$ kilowatts. Regarding $vCPU_{min_\lightning}$ and $vCPU_{max_\lightning}$, these depend on the model of the physical CPU being used. GitHub-hosted runners use Microsoft Azure's \textit{Dadsv5-series} machines~\cite{GithubRunners-2024}. The \textit{Dadsv5 series} uses AMD's third-generation EPYC 7763v processors~\cite{Microsoft_2024}. The Cloud Carbon Footprint methodology uses data from the SPECpower Committee~\cite{SPECPower_2024} to calculate $vCPU_{min_\lightning}$ and $vCPU_{max_\lightning}$ for each CPU architecture group. For third-generation EPYC processors, the \textit{ccf-coefficients} tool\footnote{\label{footnote:ccfcoefficients}\url{https://github.com/cloud-carbon-footprint/ccf-coefficients}} used by Cloud Carbon Footprint calculates the values \SI{4.34e-4}{\kilo\watt} and \SI{1.948e-3}{\kilo\watt}, respectively, for $vCPU_{min_\lightning}$ and $vCPU_{max_\lightning}$.
\\
\subparagraph{$\mathbf{Storage_\lightning}$.} Equation~\ref{eq:storage} shows the formula to calculate the energy spent using storage units. Each GitHub-hosted runner reserves 14GB of SSD storage~\cite{GithubRunners-2024} (\textit{Reserved Storage}). For SSD storage, the Cloud Carbon Footprint methodology uses a storage coefficient of $0.0012\ kWh$ per Terabyte-Hour (\textit{Storage Coefficient}). The storage coefficient is calculated according to the 2016 US Data Center Usage Report~\cite{shehabi2016united}. To validate this coefficient, we compared it with sustainability data from multiple Seagate SSD models~\cite{Seagate2020,Seagate2020b,Seagate2020c,Seagate2020d}, finding a close alignment between the reported figures and the Cloud Carbon Footprint estimate.
\\
\subparagraph{$\mathbf{Memory_\lightning}$.} Equation~\ref{eq:memory} states that the energy spent on RAM depends on the chosen memory coefficient (\textit{Memory Coefficient}) and on the average memory usage ($\overline{\textit{Memory}_\textit{usage}}$). The memory coefficient was set to $0.000392 kWh$ per Gigabyte-Hour. The Cloud Carbon Footprint team calculated this value by averaging the values provided by memory manufacturers~\cite{CloudCarbonFootprint-2024}.
\\
\subparagraph{$\mathbf{Network_\lightning}$.} There is an energy cost associated with running the network infrastructure required to download and upload data. Equation~\ref{eq:network} describes how we calculate that cost. Given that data centers tend to have very efficient networks, the Cloud Carbon Footprint team chose the most conservative estimate available to date of 0.001 kWh per GB of transferred data~\cite{CloudCarbonFootprint-2024} (\textit{Network Coefficient}).\\

\paragraph{Embodied Emissions} The Cloud Carbon Footprint methodology follows the formula for embodied emissions specified in the Software Carbon Intensity Specification of the Green Software Foundation~\cite{Green_Software_Foundation_Software_Carbon_Intensity_2024}:

\begin{align}
    &\textit{Embodied Emissions} =\nonumber\\
&\quad\begin{aligned}
& \textit{Total Embodied Emissions}~[MTCO2e]\ \times\\
& \frac{\textit{Exec Time}~[s]}{\textit{Expected Lifespan}~ [s]}\ \times\ \frac{\textit{Reserved Resources}}{\textit{Total Resources}}
\end{aligned}
\label{eq:embodied}
\end{align}

In our case, the total embodied emissions (\textit{Total\ Embodied\ Emissions}) are the sum of the life cycle assessment (LCA) emissions for all hardware components~\cite{CloudCarbonFootprint-2024} used to execute the GitHub Actions pipelines. We used the \textit{ccf-coefficients}\footnote{See footnote~\ref{footnote:ccfcoefficients}.} tool to calculate the total embodied emissions for Microsoft Azure’s Dadsv5-series machines, which are the machines used in GitHub-hosted runners. To calculate total embodied emissions, the \textit{ccf-coefficients} tool follows a methodology presented by Davy~\cite{Davy_2021}. We obtain a value of $1.61079\ MTCO2e$ for the total embodied emissions.
The expected lifespan (\textit{Expected\ Lifespan}) is the expected time that the equipment will remain installed~\cite{CloudCarbonFootprint-2024}. The Cloud Carbon Footprint team sets this value at 4 years~\cite{CloudCarbonFootprint-2024}, based on the Dell PowerEdge R740 Full Life Cycle Assessment~\cite{Busa_2019}.

The proportion between reserved resources and total resources is given by dividing the number of reserved vCPUs (\textit{Reserved\ Resources}) by the maximum number of vCPUs on the bare metal machine used (\textit{Total\ Resources}). The Cloud Carbon Footprint methodology sets \textit{Total\ Resources} as the number of vCPUs in the largest instance of the given family of instances~\cite{CloudCarbonFootprint-2024}. The largest instance in the \textit{Dadsv5-series} machines has 96 vCPUs~\cite{Microsoft_2024}. However, Microsoft offers a dedicated host with the same AMD third-generation EPYC 7763v processor~\cite{Microsoft_Eadsv_2024}, which provides 112 vCPUs. Therefore, we set \textit{Total\ Resources} to this value. We set \textit{Reserved\ Resources} at 4, since that is the number of vCPUs used for the GitHub-hosted runners~\cite{GithubRunners-2024}.

\subsubsection{Estimating the Average Carbon Footprint of a Repository}
\label{sec:estimatingaverage}

\begin{figure}
    \centering
    \includegraphics[width=0.8\linewidth]{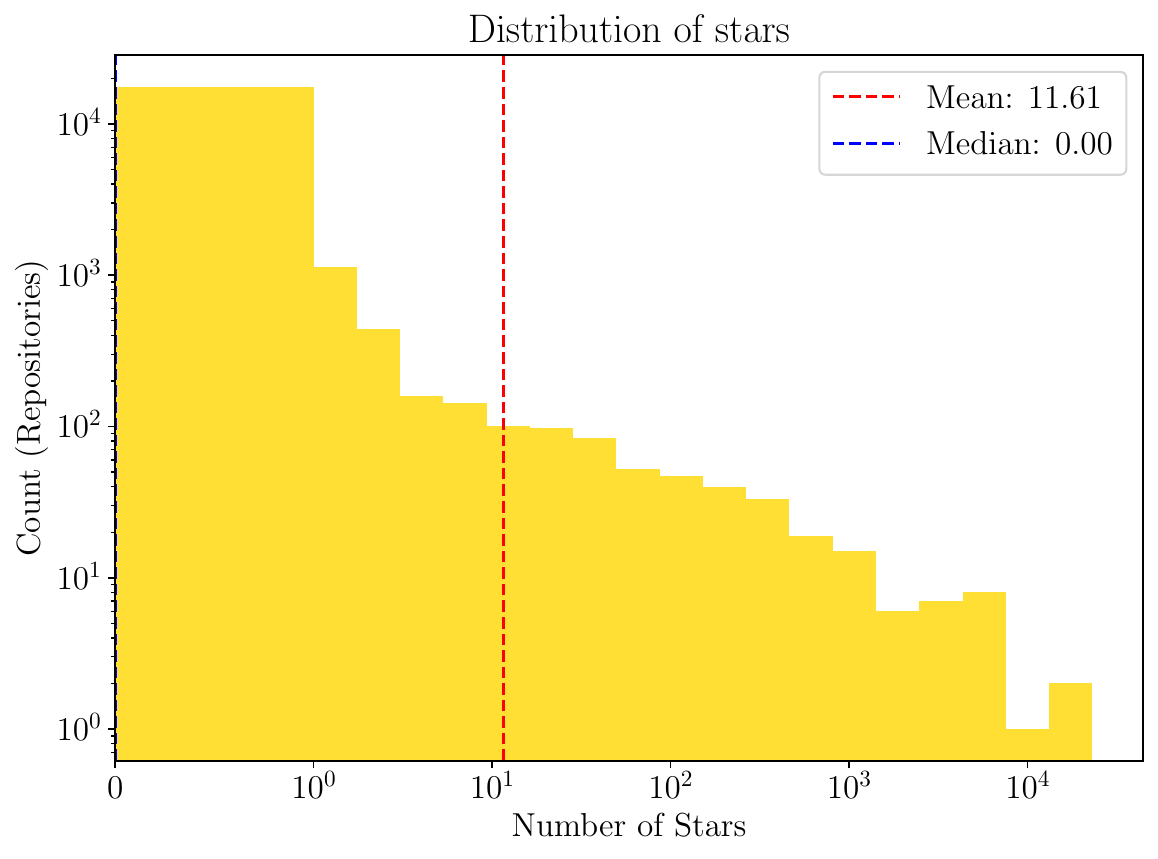}
    \caption{Distribution of stars for repositories in our 2024 GitHub Actions sample. The y-axis uses a logarithmic scale, while the x-axis combines a linear scale up to 1 and a logarithmic scale beyond that point.}
    \label{fig:repo_stars}
\end{figure}

\begin{figure}
    \centering
    \includegraphics[width=0.8\linewidth]{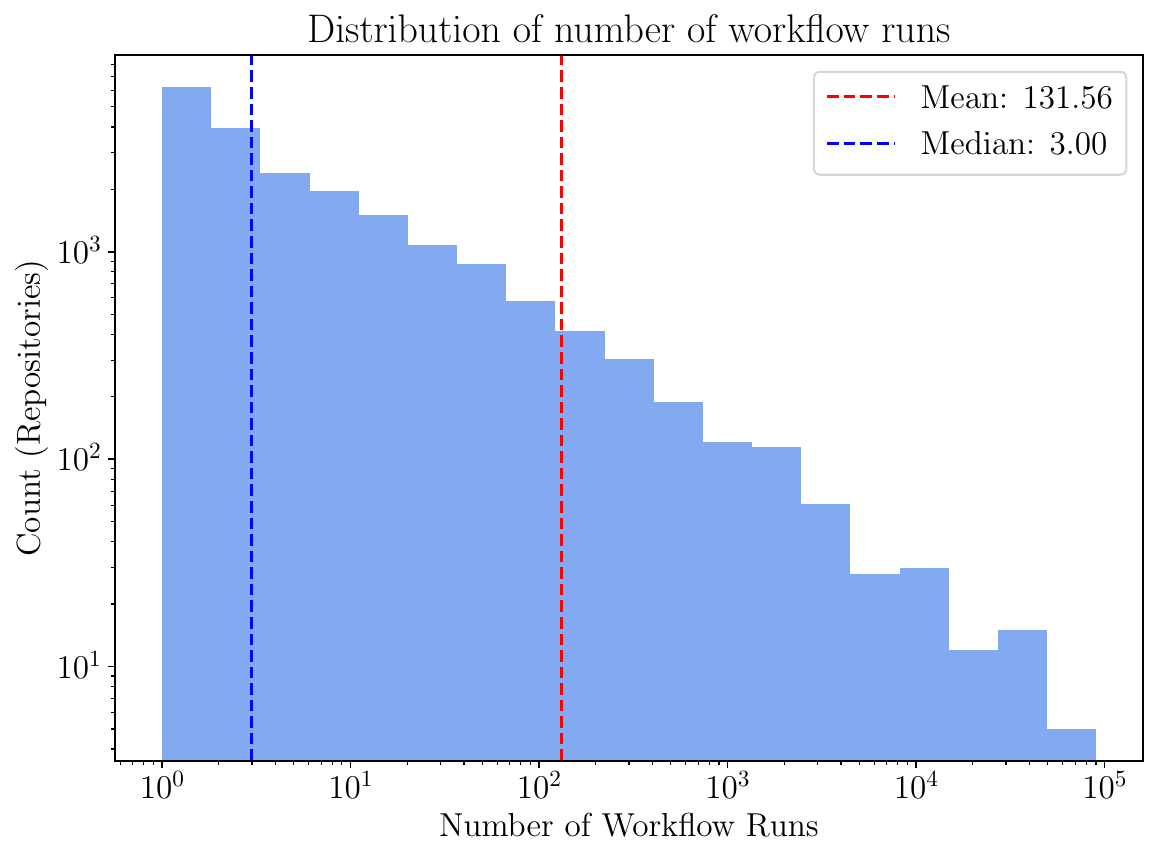}
    \caption{Distribution of number of workflow runs per repository in our 2024 GitHub Actions sample. Both the x-axis and y-axis are in logarithmic scale.}
    \label{fig:workflow_runs}
\end{figure}

To calculate the average carbon footprint of an active repository using GitHub Actions ($\overline{C_f}$), we use the GitHub API to gather workflow run data for the 20,001 repositories actively using GitHub Actions collected in Section~\ref{sec:active-repos}.
We successfully collected \workflowRuns workflow runs that include 3,446,572 jobs from \repositories repositories. Figures~\ref{fig:repo_stars} and \ref{fig:workflow_runs} respectively show the distribution of stars and the number of workflow runs per repository in our sample. We were unable to retrieve workflow run data for the remaining 1,318 repositories due to GitHub API errors, such as forbidden access errors. We filter out jobs where the completion date precedes the start date, which may occur when jobs are instantly skipped, fail immediately, or due to unexpected bugs. We also filter out jobs that lack a completion date and those with a duration greater than seven hours. Since GitHub Actions enforces a six-hour job limit, we include an additional hour as a buffer to account for possible completion delays.

Then, we calculate the carbon footprint of each repository in our sample. To calculate the carbon footprint of each repository, we sum the carbon footprint of each job in the workflow runs collected for the given repository. We apply Equation~\ref{eq:total_co2} to calculate the carbon footprint of each job.

\begin{figure}
    \centering
    \includegraphics[width=\linewidth]{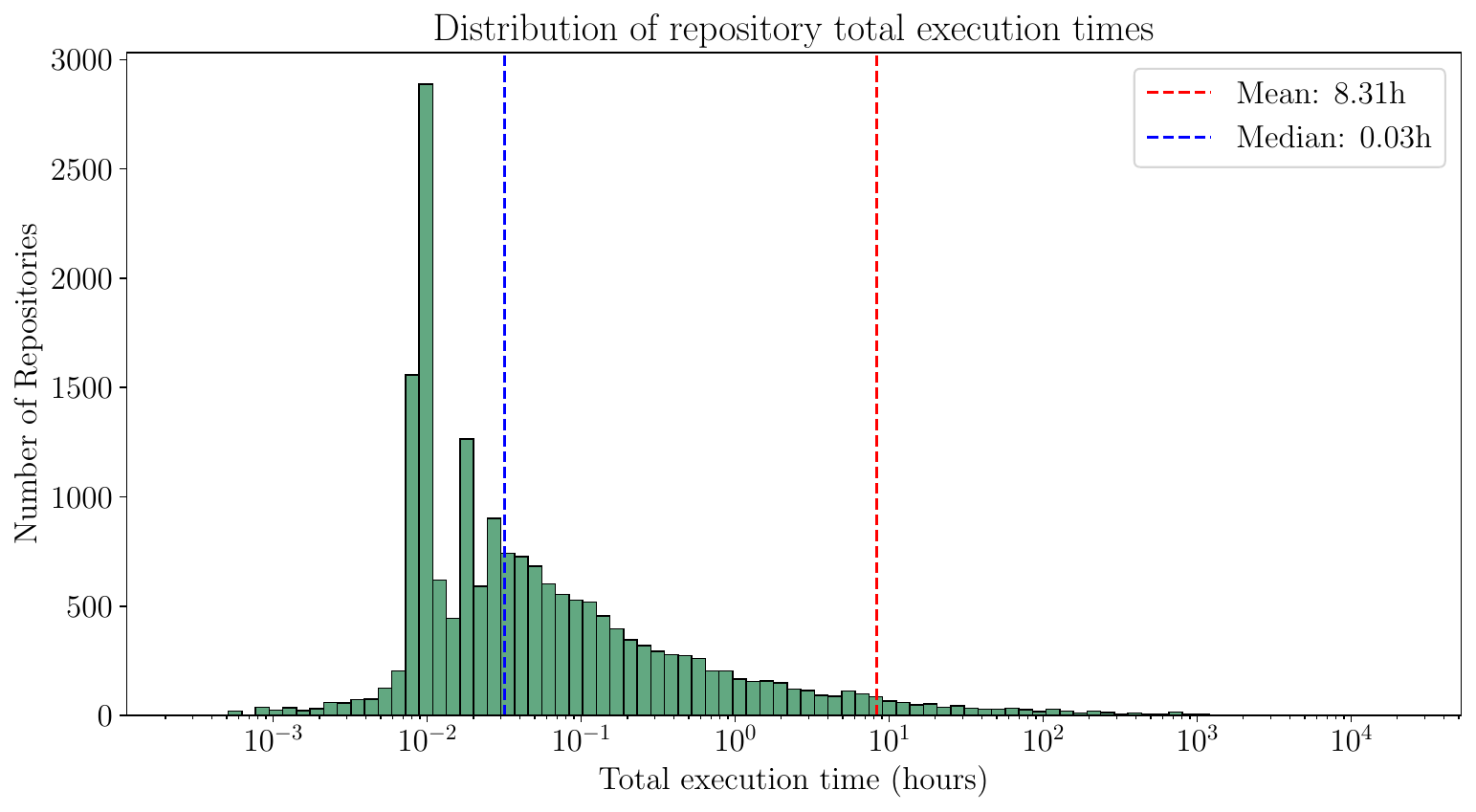}
    \caption{Distribution of total execution times for repositories in our 2024 GitHub Actions sample. The x-axis is in logarithmic scale.}
    \label{fig:execution_time_distribution}
\end{figure}

In Section~\ref{sec:cloud-carbon-footprint}, we define the following values that depend on the execution profile of the GitHub Actions ecosystem:
\begin{enumerate}[itemsep=5pt]
    \begin{multicols}{2}
        \item \textit{Exec\ Time}
        \item $\overline{vCPU_\textit{usage}}$
        \item $\overline{\textit{Memory}_\textit{usage}}$
        \item $\textit{Network}_\textit{usage}$
    \end{multicols}
\end{enumerate}
The execution time (\textit{Exec\ Time}) for each job is provided by the GitHub API.
Figure~\ref{fig:execution_time_distribution} shows the distribution of the total execution time per repository in our sample dataset of \workflowRuns workflow runs.

However, the GitHub API does not provide any information about CPU, memory, or network usage. For that reason, we must estimate the values for these metrics.

\paragraph{Execution Metrics}\label{sec:exec_metrics} To calculate $\overline{vCPU_\textit{usage}}$, $\overline{\textit{Memory}_\textit{usage}}$, and $\textit{Network}_\textit{usage}$, we must re-run the workflow runs while monitoring these metrics. To do so, we use the same sample of repositories as in Section~\ref{sec:estimatingaverage}. Then, we filter out the repositories that do not have at least a workflow run that fulfills the following criteria:
\begin{enumerate}
    \item The workflow run completed successfully. If we considered workflow runs that failed, we would not be able to ensure that the run was reproduced successfully or if it failed for a different reason from the original run.
    \item The workflow run does not use secrets. In GitHub Actions, secrets are variables that you create in an organization, repository, or repository environment~\cite{GithubSecrets-2024}. Since we cannot access these secrets, we would not be able to reproduce workflow runs that use them.
    \item We have access to the original workflow file. Some workflow runs do not originate from workflow files (e.g. GitHub pages deployments), or these files are no longer available. Since we manipulate these files to reproduce the run, we need them to be available.
\end{enumerate}

After filtering, 2,934 out of \repositories repositories remained. For each remaining repository, we select a workflow run that fulfills the above criteria. Then, we clone the repository and push a replica to GitHub. In this replica, we instrument the run workflow so that we can manually trigger it. Furthermore, we add a step to the beginning of each job in the workflow that runs the \textit{workflow-telemetry} action\footnote{\url{https://github.com/catchpoint/workflow-telemetry-action}}. The \textit{workflow-telemetry} action collects CPU, memory, network, and other metrics throughout the run of a job. We modified the \textit{workflow-telemetry} action to obtain the raw metrics instead of the provided plots. Finally, we trigger each workflow run and collect its metrics. We only keep the data for workflow runs that complete successfully, that is, those that are reproduced successfully.
We were able to replicate a workflow run from 1,463 repositories.

Given that we are unable to reproduce all workflow runs in our sample, we estimate average values for each metric and use them to calculate the carbon footprint for all jobs. We believe that this approach provides a better estimate than only considering workflow runs that we are able to reproduce since, for instance, for more complex and longer runs, we are less likely to be able to reproduce them. For the same reason, we will consider three different settings for the metrics: one with the estimated average values (\textit{Baseline}), one where these values are doubled (\textit{High\ Usage}), and one where they are halved (\textit{Low\ Usage}). The additional two settings allow us to understand what the carbon footprint would be if the resource usage was actually higher or lower than our estimate.
\\
\subparagraph{CPU usage.} To obtain $\overline{vCPU_\textit{usage}}$, we calculate the weighted average of vCPU usage with respect to execution time. Since the \textit{workflow-telemetry} action collects metrics at fixed intervals of time, we sum all the values collected and divide by the number of samples. Finally, as the \textit{workflow-telemetry} action collects the vCPU usage as the percentage of total load, we multiply this percentage by $4$ (the number of available vCPUs) to obtain $\overline{vCPU_\textit{usage}}$, which resulted in a value of 1.51 vCPUs.
\\
\subparagraph{Memory usage.} As for the vCPU usage, we obtain the $\overline{\textit{Memory}_\textit{usage}}$ by summing all the values collected by the \textit{workflow-telemetry} action and dividing by the number of samples, which resulted in a value of 1.78 GB.
\\
\subparagraph{Network usage.} In our case, $\textit{Network}_\textit{usage}$ corresponds to the average network usage per job. To calculate it, we sum the network usage of all jobs and divide it by the number of total jobs in the collected workflow runs, which resulted in a value of 0.22 GB.
\\
\\
Finally, $\overline{C_f}$ is calculated by adding the carbon footprint of all the repositories in our sample and dividing by the size of the sample. In our study, this value ranges between $1.36\text{e-}5$ MTCO2e and $5.86\text{e-}5$ MTCO2e depending on the region we are considering.

\subsubsection{Simplification Assumptions}

To simplify our estimates, we apply the following assumptions.
\begin{enumerate}
    \item We consider that macOS runners run on the same type of machine as Windows and Linux. Depending on the macOS version used, these machines can have fewer CPUs and memory than Windows and Linux machines~\cite{GithubRunners-2024}. Furthermore, even though these machines are hosted in Azure data centers, they do not run on the Microsoft Azure service and we have no guarantees that they are similar to the Dadsv5-series machines~\cite{GithubRunners-2024}. Only 1.7\% of the runs in our data use macOS runners.
    \item Similarly to macOS runners, we assume self-hosted runners to have the same specification as Windows and Linux GitHub-hosted runners. Only 0.8\% of the runs in our data use self-hosted runners.
    \item We ignore the energy consumption by GPUs since standard and free GitHub-hosted runners do not provide access to GPUs. Some self-hosted and paid GitHub-hosted runners may use GPUs and their usage might be significant, but the majority of repositories in the GitHub Actions ecosystem only use CPUs for computation.
    \item According to the Cloud Carbon Footprint team, the electricity used to power data exchange inside the same data center is close to 0. Given that we do not know the source and destination of the data transferred, we assume that all data is transferred between different data centers. The 0.001 kWh per GB of transferred data is a conservative estimate, and so we believe that even with this assumption, our approach might still underestimate the real value.
\end{enumerate}

\subsection{Water footprint}

To calculate the average water footprint of a repository ($\overline{W_f}$), we follow a similar approach to that presented by Jiang et al.~\cite{jiang2025waterwise}. We use the same sample of workflow runs as in Section~\ref{sec:average-carbon}. According to Jiang et al's approach, the total water footprint is composed of the sum of three components as follows.

\begin{align}
    & \textit{Water\ Footprint} =\nonumber \\
    & \qquad\begin{aligned}[t]
        &\phantom{~+~}\textit{Operational\ Water\ Footprint}_\textit{offsite} \\
        &+\ \textit{Operational\ Water\ Footprint}_\textit{onsite} \\
        &+\ \textit{Embodied\ Water\ Footprint}
    \end{aligned}
\end{align}

The $\textit{Operational\ Water\ Footprint}_\textit{offsite}$ refers to the water consumed during the production of electricity that powers a data center and is calculated as follows.

\begin{align}
    & \textit{Operational\ Water\ Footprint}_\textit{offsite} =\nonumber \\
    & \qquad\begin{aligned}[t]
        & \phantom{~\times~}\textit{Cloud\ Energy\ Consumption}~[kWh]\\
        & \times\ \textit{PUE}\ \times\ \textit{EWIF}\ \Bigr[\frac{L}{kWh}\Bigr]
    \end{aligned}
\end{align}

The Energy Water Intensity Factor (\textit{EWIF}) quantifies the amount of water consumed to produce a unit of electricity~\cite{jiang2025waterwise}. This factor is highly dependent on the energy sources that comprise the regional energy mix, as different sources have varying water consumption profiles. In our work, we adopt the grid-average water use factors reported by Reig et al.~\cite{reig2020guidance} for each of the regions under consideration.

The $\textit{Operational\ Water\ Footprint}_\textit{onsite}$ refers to the water consumed in the data center for cooling purposes, and is calculated as follows.

\begin{align}
    & \textit{Operational\ Water\ Footprint}_\textit{onsite} =\nonumber \\
    & \qquad\begin{aligned}[t]
        & \phantom{~\times~}\textit{Cloud\ Energy\ Consumption}~[kWh]\\
        & \times\ \textit{WUE}\ \Bigr[\frac{L}{kWh}\Bigr]
    \end{aligned}
\end{align}

The water usage effectiveness (\textit{WUE}) of a data center quantifies the amount of water required to dissipate heat per unit of energy consumed~\cite{jiang2025waterwise}. \textit{WUE} varies with the geographical location of the data center, --- cooler climates generally require less water --- and the efficiency of the cooling systems in place.
We use the region-specific values provided by the Microsoft Sustainability Team: 0.55 for the Americas, 1.65 for Asia Pacific, and 0.1 for Europe, the Middle East, and Africa~\cite{Noelle2022}.

Lastly, similar to the embodied emissions, the \textit{Embodied\ Water\ Footprint} represents, in our case, the water consumed during the manufacturing of the hardware components used to execute the GitHub Actions pipelines. Due to the lack of publicly available data on the embodied water footprint, Jiang et al.~\cite{jiang2025waterwise} propose an estimation based on the corresponding embodied carbon footprint.

\begin{align}
    & \textit{Embodied\ Water\ Footprint} =\nonumber \\
    & \qquad\begin{aligned}[t]
        & \phantom{~\times~}\textit{E}_\textit{manufacturing}~[kWh]\\
        & \times\ \textit{EWIF}_\textit{manufacturing}~\Bigr[\frac{L}{kWh}\Bigr]
    \end{aligned}
\end{align}

The method estimates the energy used in manufacturing ($\textit{E}_\textit{manufacturing}$) by dividing the carbon footprint by the carbon intensity of the region where the hardware was manufactured. This estimated energy consumption is then multiplied by the manufacturing region's \textit{EWIF} to obtain the embodied water footprint.

Since the manufacturing locations of Azure's servers are not publicly disclosed, we estimated the impact using a weighted average based on the major semiconductor-producing countries. We follow Davy’s findings~\cite{Davy_2021} that indicate that the majority of the embodied carbon footprint of a server comes from semiconductor production, and we apply the same assumption to estimate the embodied water footprint.

According to a report by the United States Congress, the United States, Taiwan, South Korea, Japan, and China are identified as the countries with the largest semiconductor manufacturing capacities, accounting for approximately 10\%, 18\%, 16\%, 17\%, and 22\% of global capacity, respectively~\cite{Blevins2023}. We used these values to compute a weighted average of the carbon intensity and \textit{EWIF} in these five regions.

Since server manufacturing dates are unknown, we adopt static values for carbon intensity in our analysis. For each country considered, we used the average values of carbon intensity for 2024 provided by Electricity Maps~\cite{ElectricityMapsYearly2025}. For \textit{EWIF}, we used the grid-average water use factors reported by Reig et al.~\cite{reig2020guidance}, with the exception of Taiwan, for which such data were not available, and therefore we used the values provided by Chen et al.~\cite{chen2015quantifying}.

\subsection{Regions}
\label{sec:regions}

In our study, we included all Azure regions~\cite{Microsoft2025} and their respective countries for which data on carbon intensity, energy-water intensity factors, or both were available. Considering country-level data provides insight subparagraphinto emissions and water use for regions where more granular region-specific values are unavailable.

At the time of writing, Azure was operating in 54 regions in 30 countries. Among these regions, carbon intensity data were available for 29, and energy-water intensity factor data were available for 28. At the country level, carbon intensity data were available for all 30 countries, while energy-water intensity factor data were available for 25.

For our most likely scenario, we use average values derived from the five US regions where runners are deployed according to the GitHub support~\cite{GithubActionsRegions-2021}.

\section{Results}
In this section, we present and discuss the results of following the methodology described in Section~\ref{sec:cloud-carbon-footprint}.
To address RQ1, we begin by presenting the CWF of the GitHub Actions ecosystem for the year 2024. Subsequently, in response to RQ2, we propose a set of strategies to reduce this footprint.

\subsection{Carbon Footprint (RQ1)}
\label{sec:results_yearly_carbon_footprint}

\begin{figure*}
    \centering
    \includegraphics[width=\linewidth]{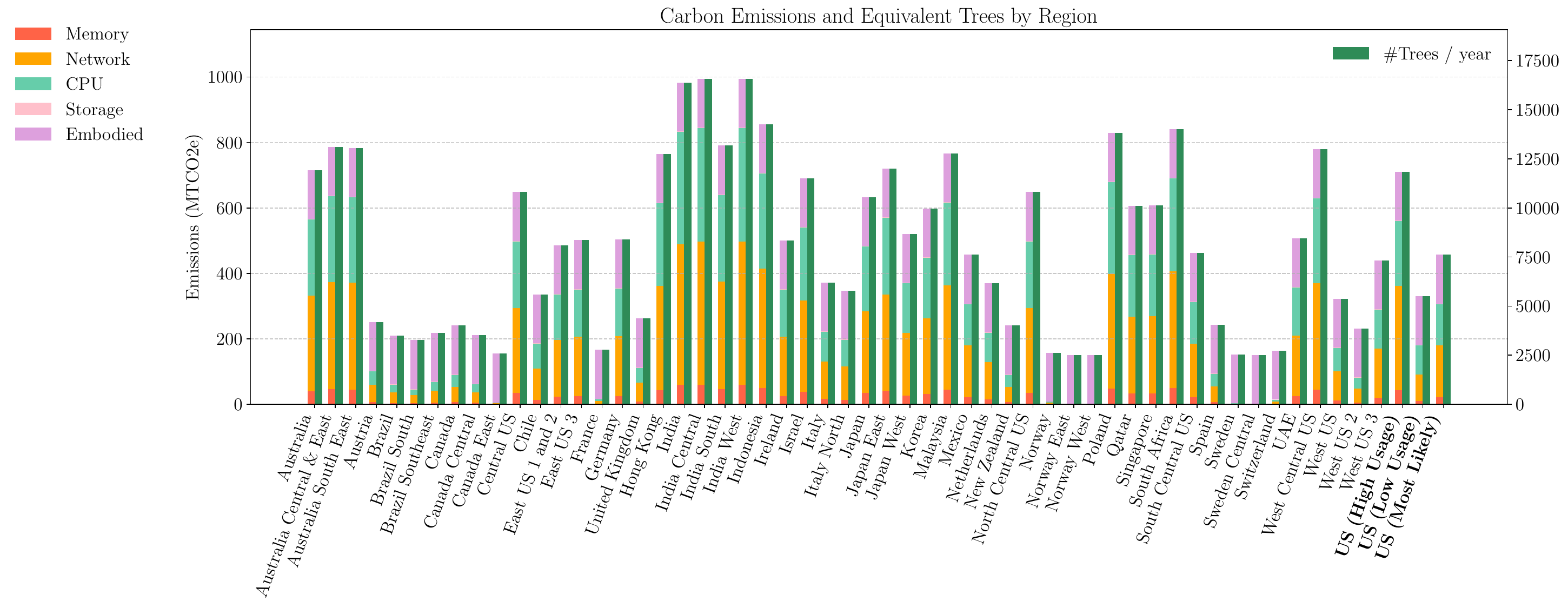}
    \caption{Carbon emissions of the GitHub Actions ecosystem in 2024 depending on the region where the runners are deployed (left y-axis). Number of urban tree seedlings necessary to capture those carbon emissions in a year~\cite{EPA-2024} (right y-axis).}
    \label{fig:region_emissions}
\end{figure*}

\begin{figure*}
    \centering
    \includegraphics[width=\linewidth]{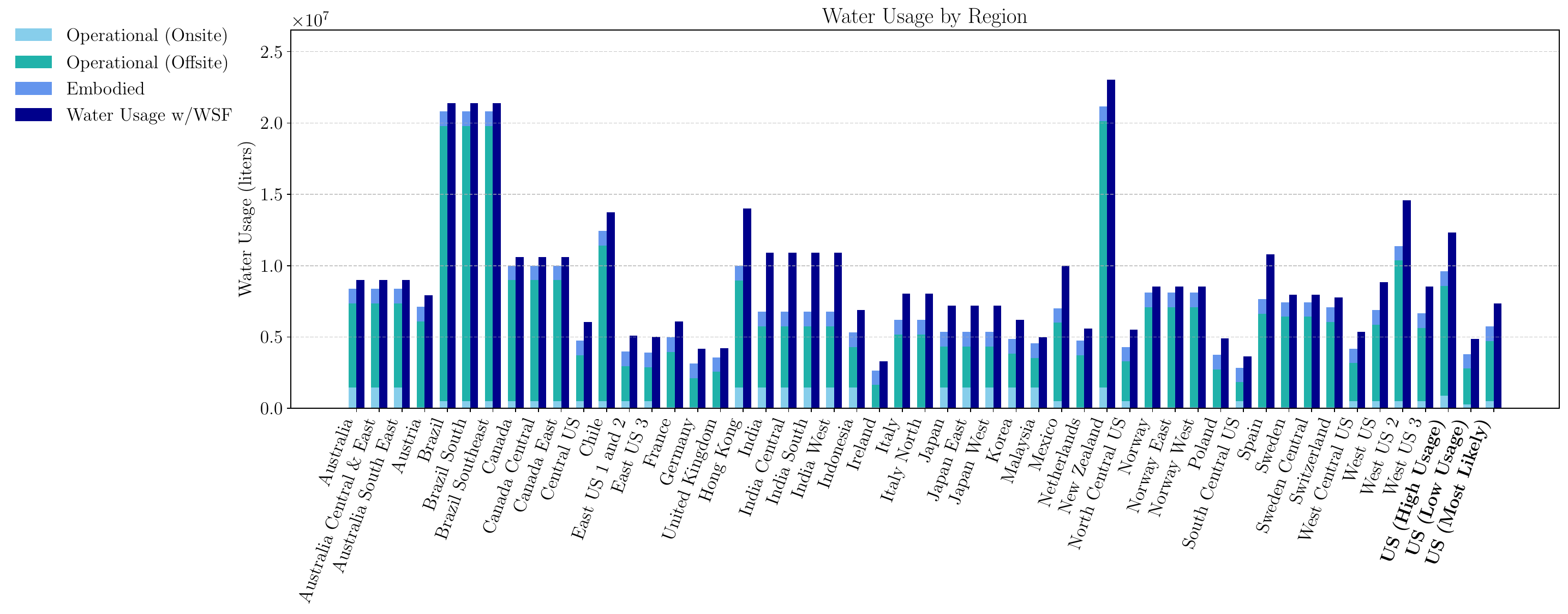}
    \caption{Water usage of the GitHub Actions ecosystem in 2024 depending on the region where the runners are deployed. In dark blue, the plot shows the relative water usage across regions adjusted by the water scarcity factor.}
    \label{fig:region_water}
\end{figure*}

Figure \ref{fig:region_emissions} shows the results of the yearly estimate of the carbon footprint for each region in Section~\ref{sec:regions}. Our yearly estimate varies between  \emissionsOptimisticScenario and \emissionsPessimisticScenario MTCO2e, depending on the region.
For our most likely region, we also consider the three different settings defined in Section~\ref{sec:estimatingaverage}.

The region with the lowest carbon emissions is \textit{Norway West}, due to its exceptionally low grid emission factor. Other regions in Norway, as well as in \textit{Sweden}, \textit{France}, and \textit{Switzerland}, exhibit similarly low emission values.
The regions with the highest carbon emissions are when runners are deployed in \textit{India Central} and \textit{India West}, which have very high grid emission factors.

As for our most likely scenario, with the \textit{Baseline} setting the 2024 carbon footprint of the GitHub Actions ecosystem would be \emissionsLikelyScenario MTCO2e.
If we consider the \textit{Low\ Usage} setting, the value decreases to 330.5 MTCO2e, while for the \textit{High\ Usage} setting, the value increases to 709.7 MTCO2e.

Figure \ref{fig:region_emissions} also shows the number of urban tree seedlings required to capture the carbon emitted in 2024 by the GitHub Actions ecosystem. We calculate this value according to the estimate by the United States Environmental Protection Agency (EPA) that an urban tree seedling allowed to grow for 10 years captures on average 0.060 MTCO2e per year~\cite{EPA-2024}. For our most likely scenario, to capture all the carbon emitted in 2024 by the GitHub Actions ecosystem, \num{7615} urban tree seedlings would be required.

Figure~\ref{fig:emissions_comparison} compares the yearly carbon emissions of the GitHub Actions ecosystem in our most likely scenario to the emissions of quotidian activities. We note that for the calculations involving the equivalent amounts of fried chicken prepared in an air fryer and fully charged smartphones, we use the average of the yearly average grid emission factors across the five regions considered in our most likely scenario.

\subsection{Water Footprint (RQ1)}
\label{sec:results_yearly_water_footprint}

Figure~\ref{fig:region_water} shows the yearly water footprint of the GitHub Actions ecosystem for each region in Section~\ref{sec:regions}. As in Section~\ref{sec:results_yearly_carbon_footprint}, for our most likely region we consider the three different settings defined in Section~\ref{sec:estimatingaverage}. Our yearly estimate varies between \waterOptimisticScenario and \waterPessimisticScenario kiloliters.

The region with the lowest water usage is \textit{Ireland}, with \textit{Germany} and the \textit{South\ Central\ United\ States} also demonstrating similarly low levels. In contrast, \textit{New\ Zealand} exhibits the highest water usage, followed closely by \textit{Brazil}.

For our most likely scenario, with the \textit{Baseline} setting, the water footprint of 2024 would be \waterLikelyScenario kiloliters. However, if we consider the \textit{Low\ Usage} setting, the value decreases to 3,802.1 kiloliters, and increases with the \textit{High\ Usage} setting to 9,610.5 kiloliters.

Figure~\ref{fig:region_water} also illustrates the relative water usage across regions when adjusted by the water scarcity factor (\textit{WSF}). Countries with greater water availability have a lower \textit{WSF}, while countries with more limited water resources exhibit a higher \textit{WSF}. We multiply the water usage by \textit{WSF} as in the work of Jiang et al.~\cite{jiang2025waterwise}. For example, when \textit{WSF} is taken into account, \textit{Spain}, despite having a lower absolute water usage than \textit{Norway}, exhibits a significantly higher relative water usage.

Figure~\ref{fig:water_comparison} compares the yearly water usage of the GitHub Actions ecosystem in our most likely scenario to the water usage of quotidian activities.

\subsection{Reducing the CWF of the GitHub Actions ecosystem (RQ2)}

In this section, we analyze the data collected in Sections~\ref{sec:results_yearly_carbon_footprint} and \ref{sec:results_yearly_water_footprint} to suggest strategies to reduce the CWF of the GitHub Actions ecosystem.

\subsubsection{Environment-aware region selection}

Figures~\ref{fig:region_emissions} and ~\ref{fig:region_water} illustrate that the carbon and water footprint of the GitHub Actions ecosystem can vary significantly depending on the region where the runners are deployed.

The carbon footprint can be reduced by up to 67.1\% (\textit{Norway\ West}) or increased by up to 217.7\% (\textit{India\ Central} and \textit{India\ West}) relative to our most likely scenario. In \textit{Norway}, \textit{Sweden}, \textit{France}, \textit{Switzerland}, and \textit{Canada\ East} the grid emission factor is so low that operational emissions are negligible compared to embodied emissions.
Claßen et al. also showed that better aligning the execution of CI/CD runs with the regional availability of low
carbon energy can reduce the carbon footprint by up to 25.3\%~\cite{classen2023carbon}.

As for the water footprint, depending on the region it can be reduced by up to 65.3\% (\textit{Ireland}) or increased by up to 556.4\% (\textit{New\ Zealand}) relative to our most likely scenario.

There is a notable trade-off between the carbon and water footprint in some regions. For example, regions such as \textit{Brazil} exhibit relatively low carbon emissions, yet incur a substantially high water footprint. In contrast, regions such as \textit{India} demonstrate high carbon emissions while maintaining a comparatively low water footprint. This trade-off is related to the energy mix used in each region.

The energy mix plays a crucial role in shaping the environmental footprint of a region. Regions that heavily depend on hydroelectric or geothermal power typically have higher water footprints, as these energy sources consume more water per kilowatt hour generated. For example, according to data presented by Reig et al.~\cite{reig2020guidance}, hydroelectric power accounted for 63.3\% of Brazil’s energy mix, with each kilowatt-hour requiring approximately 27 liters of water. In contrast, India derives 59.3\% of its energy from hard coal, which consumes only about 2 liters of water per kilowatt-hour. However, this lower water intensity comes at the cost of significantly higher carbon emissions, since hard coal is a non-renewable, carbon-intensive energy source, unlike hydro, which is renewable in terms of carbon impact.

\begin{tcolorbox}[colback=yellow!20!white,colframe=yellow!70!black,arc=10pt]
\textbf{Call to action:} To reduce the environmental impact, GitHub could choose the regions where runners are deployed considering the region's grid emission factors and water consumption. Regions such as \textit{France} and \textit{United\ Kingdom} should be preferred as they present a good trade-off between carbon and water footprints. Moreover, GitHub should explicitly display to users the region where each runner is deployed and the region's environmental performance.
\end{tcolorbox}

\subsubsection{Scheduled runs}
\label{sec:scheduled_runs}

Workflow files enable developers to define the execution of workflow runs at specific times or on a recurring schedule. These runs are called scheduled runs.
In their dataset of workflow runs, Bouzenia et al. identified that 15.5\% of the overall execution time was consumed by scheduled runs~\cite{bouzenia2024resource}. However, among the \workflowRuns workflow runs we analyzed, approximately 33.9\% of the total execution time was attributed to scheduled runs.

In addition to geographical load shift, Claßen et al. demonstrated that carbon intensity–aware temporal shifting of CI/CD workloads by up to six hours can achieve an additional reduction in carbon emissions of approximately 6\%. For our dataset, we simulated the impact of deferring scheduled CI/CD runs to the hour of lowest carbon intensity within the same calendar day on which the original execution occurred.
Unlike other CI/CD workflow runs, scheduled runs typically do not require immediate or fast feedback to developers. As a result, they offer greater flexibility for alignment with periods of lower carbon intensity on the energy grid. In our most likely scenario, we found that such a temporal alignment could reduce carbon emissions by up to 3.9\%.
A comparable reduction in water usage can also be expected if scheduled runs align with periods when the regional energy mix is optimized to minimize both carbon and water footprints.

Moreover, of the total execution time for scheduled runs, around 10.9\% of the time was in inactive repositories. We consider a repository to be inactive if a job was executed more than 30 days after the last push and no subsequent push has occurred by our collection date. All repositories had no pushes for at least 36 days, with a median period of 305 days, which allows us to confirm their inactivity with greater certainty.

An example of an inactive repository wasting computation time through scheduled runs was \textit{AstaTus/openssl}\footnote{\url{https://github.com/AstaTus/openssl}}. \textit{AstaTus/openssl} is a fork of the \textit{openssl/openssl}\footnote{\url{https://github.com/openssl/openssl}} repository. Although scheduled runs are disabled by default in forks of public repositories~\cite{GitHubDisablingWorkflows_2024}, users can activate them. The scheduled runs of this repository consumed 626 hours of computation in 2024 before being halted due to GitHub's policy, which disables scheduled runs after 60 days of inactivity in the repository~\cite{GitHubDisablingWorkflows_2024}. During our analysis, we found other similar examples to \textit{AstaTus/openssl}.

Workflow runs on forks consumed around 37.7\% of the total execution time of all considered runs. Of those 37.7\%, 61.5\% were consumed by scheduled runs, which is 27.6\% higher than when we consider all repositories.

\begin{tcolorbox}[colback=yellow!20!white,colframe=yellow!70!black,arc=10pt]
\textbf{Call to action:} Both GitHub and developers should be careful when using scheduled runs, particularly on forks. GitHub's policies of disabling scheduled runs by default on forks and after 60 days of repository inactivity are a commendable step in this direction. However, our findings suggest that additional strategies, such as those proposed by Bouzenia et al.~\cite{bouzenia2024resource}, could further address these issues. The authors recommend deactivating scheduled workflows after $k$ consecutive failures and imposing stricter criteria for deactivating scheduled runs due to inactivity. The execution of scheduled runs could also be timed to coincide with periods when the regional energy mix is more favorable to minimize environmental impact.
\end{tcolorbox}

\subsubsection{Disclosing the CWF}

Studies indicate that informing users about the carbon footprint of their activities can reduce their carbon emissions by up to 35\%, depending on the type of activity~\cite{hoffmann2024carbon}. Studies have also shown that people feel more guilty about their carbon emissions when they learn that they, or a group to which they belong, create more carbon emissions than their peers~\cite{mallett2013self}.

\begin{tcolorbox}[colback=yellow!20!white,colframe=yellow!70!black,arc=10pt]
\textbf{Call to action:} GitHub's interface could show developers the CWF of workflow runs they trigger and those from repositories they contribute to. Moreover, GitHub could also show comparisons with the carbon and water footprints of other developers and repositories. Possible metrics include the median and average carbon and water footprints:
\begin{inparaenum}
    \item per user/repository;
    \item per workflow run;
    \item per minute of execution time.
\end{inparaenum}
\end{tcolorbox}

\subsubsection{Repository Size}

In our most likely scenario, network-related emissions represent approximately 34.8\% of total carbon emissions, constituting the largest contributing component. Meanwhile, the action \emph{checkout}\footnote{\url{https://github.com/actions/checkout}} is responsible for 12.2\% of the total execution time of our dataset. This action is responsible for retrieving the contents of the current GitHub repository onto the runner. Based on this observation, we hypothesized that a significant proportion of network-related emissions is attributable to the action \emph{checkout}.

To test this hypothesis, we cloned each repository in our dataset that contained workflows using the \emph{checkout} action. By default, the \emph{checkout} action clones repositories with a depth of 1, meaning that only the content of the last commit is retrieved. We adopt this default setting for all repositories in our dataset. Furthermore, to estimate the download size, we record the size reported by \emph{git} during the cloning process, as this reflects the compressed content and provides a more accurate measurement than the final uncompressed size on disk. It is important to note that this measurement is an approximation, as we clone the latest commit available at the time of data collection, which may differ from the specific commit used during each individual workflow run. This experiment was conducted on 19 July 2025.

We successfully collected the size of the repository for 5,529 of the 6,253 repositories that use the \emph{checkout} action in our dataset. The remaining repositories could not be cloned due to errors during cloning. The 5,529 repositories account for 1,823,639 of the 2,107,125 total checkout executions present in our dataset.
For the 1,823,639 checkout executions with available size data, we observed a cumulative downloaded size of 42,347.3 GB, yielding an average of approximately 0.02 GB per checkout. Extrapolating this average to the full set of 2,107,125 checkout executions in our dataset, we estimate a total downloaded size of 48,930.2 GB, which corresponds to about 8.1\% of our total estimated network usage.

Assuming that the \emph{checkout} action exhibits CPU and memory usage patterns consistent with our estimated averages, and accounting for both its execution time and associated network data transfer, we estimate that this action alone is responsible for approximately 6.8\% of the total carbon emissions within the GitHub Actions ecosystem.
Given that the \emph{checkout} action accounts for 12.2\% of the total execution time, we attribute an equivalent 12.2\% of the carbon emissions associated with CPU, memory, and storage usage to this action. For emissions arising from network data transfer, we attribute 8.1\% to the \emph{checkout} action, based on its estimated contribution to the overall network traffic.

\begin{tcolorbox}[colback=yellow!20!white,colframe=yellow!70!black,arc=10pt,breakable]
\textbf{Call to action:}
Our findings suggest that the \emph{checkout} action contributes approximately 6.8\% of the total carbon emissions of the platform, indicating that GitHub could reduce its environmental footprint by improving the efficiency of the cloning process. Additionally, repository maintainers are encouraged to be mindful of the repository size, as minimizing unnecessary files can help reduce resource consumption.
\end{tcolorbox}

\section{Threats to Validity}
A threat to internal validity is that, while for the number of public and active repositories and for the number of repositories using GitHub Actions we can provide confidence intervals, our estimate of the average carbon and water footprints of an active repository using GitHub Actions is not supported by statistical evidence. The three main reasons for the lack of statistical evidence are:
\begin{inparaenum}
    \item the Cloud Carbon Footprint methodology has an experimental nature that provides point estimates without confidence intervals~\cite{CloudCarbonFootprint-2024};
    \item we do not have official information about the region where GitHub Actions hosted runners are deployed;
    \item there are workflow runs that we are not able to replicate, which introduces bias in the calculation of the CPU usage, memory usage, and network usage metrics;
    \item the embodied water footprint is estimated based on the corresponding embodied carbon footprint due to lack of data;
    \item we do not have official information about the manufacturing region of the servers used by GitHub Actions.
\end{inparaenum}

To limit the impact of this threat, we calculate the carbon and water footprints using a large sample with \workflowRuns workflow runs from \repositories different repositories. To the best of our knowledge, this is the largest dataset of workflow runs in the literature. We also take into account various scenarios for the regions where the runners are deployed and for the usage metrics. This approach enables us to assess how carbon emissions might fluctuate compared to our most likely scenario.

A threat to external validity is that our research is limited to GitHub Actions and public repositories. Future work should investigate how our results compare to the carbon footprint of other CI/CD platforms, private repositories, and in industrial usage scenarios.

\section{Related Work}
\emph{GitHub Actions.} Previous research has studied the GitHub Actions ecosystem to gain insights into how developers interact with the platform, the development process of workflows and their characteristics, common issues, and the impact of the adoption of GitHub Actions~\cite{kinsman2021software,decan2022use,valenzuela2022evolution,saroar2023developers,wessel2023github,decan2023outdatedness,zhang2024developers}.
In this work, we study the GitHub Actions ecosystem to understand its carbon and water footprints.

\emph{Carbon Intensity.} Claßen et al. investigated opportunities and challenges to reduce the carbon footprint of CI/CD services by aligning their execution with periods of low-carbon energy availability, using the GitHub Actions ecosystem as a case study~\cite{classen2023carbon}. In our work, not only do we propose additional strategies to green the GitHub Actions ecosystem, but we also quantify its carbon footprint. Radovanovi{\'c} et al. apply the same type of strategy as Claßen et al. to workloads in Google datacenters~\cite{radovanovic2022carbon}.

\emph{Embodied Emissions.} Gupta et al. concluded that embodied emissions, as opposed to operational emissions, increasingly dominated the carbon footprint of mobile systems~\cite{gupta2021chasing}. Moreover, the authors mention that as more data centers employ renewable energy, the dominant source of their total
carbon footprint becomes embodied emissions~\cite{gupta2021chasing}. In our study, we observed similar findings, noting that operational emissions are negligible when GitHub Actions runners are deployed in regions with abundant green energy, such as Norway.

\emph{Network Emissions.} Zilberman et al. emphasize the critical need for carbon-efficient networking, propose potential solutions, and highlight carbon-intelligent routing as the next significant challenge in the field of networking~\cite{zilberman2023toward}. In the most likely scenario in our work, network usage is responsible for about 35\% of the carbon emissions of the GitHub Actions ecosystem, further highlighting the importance of carbon-efficient networks.

 \emph{Storage Emissions.} McAllister et al. identified three broad directions to reduce storage emissions~\cite{mcallister2024call}. The authors also mention that recent data from Azure suggest that storage-related emissions make up 33\% of operational emissions and 61\% of the embodied emissions in their data centers. According to the authors, storage will dominate overall data center emissions due to embodied storage emissions~\cite{mcallister2024call}. In our study, the operational emissions associated with storage are negligible. Regarding embodied emissions, the Cloud Carbon Footprint methodology does not provide a breakdown of embodied emissions by component, making it impossible to determine the specific percentage attributable to storage. Future research should be conducted on this topic.

\emph{Carbon Footprint methodology.} Similarly to the Cloud Carbon Footprint methodology, Simon et al. present a bottom-up methodology for assessing the environmental impacts of servers and cloud instance solutions based on crowd-sourced data~\cite{simon2024boaviztapi}. The authors argue that the Cloud Carbon Footprint methodology may not be suitable for non-computing instances, such as storage servers, and that it primarily focuses on the carbon footprint without considering other environmental impacts~\cite{simon2024boaviztapi}. In our work, since we exclusively consider computing instances and are only concerned with the carbon footprint, we use the Cloud Carbon Footprint methodology.

\emph{CI/CD optimizations.} Optimizing CI/CD runs has a direct impact on their carbon footprint. Bouzenia and Pradel describe optimization opportunities in GitHub Actions workflows~ \cite{bouzenia2024resource}. These optimizations include running previously failed jobs first, job-specific timeouts, and the optimizations related to scheduled runs mentioned in Section~\ref{sec:scheduled_runs}.
Research has also focused on identifying commits in which CI/CD runs can be safely skipped~\cite{abdalkareem2019commits,abdalkareem2020machine,saidani2021detecting}. Minimizing the number of CI/CD runs executed in each repository represents a crucial step toward reducing the carbon footprint of the GitHub Actions ecosystem.

\emph{Water footprint.} Ristic et al. conducted a preliminary study of the water footprint associated with cooling systems and energy consumption in data centers~\cite{ristic2015water}. Their findings indicate that energy consumption accounts for the vast majority of the water footprint in such facilities.
This observation aligns with our results, in which water use associated with electricity generation constitutes approximately 73.7\% of the total water footprint in our most likely scenario. A comparable value is reported in the study by Siddik et al., which estimates that approximately 75\% of the water footprint of U.S. data centers is attributable to energy consumption~\cite{siddik2021environmental}.
Wu et al. developed a framework for evaluating the water impacts of computing that incorporates spatial and temporal variations in water stress~\cite{wu2025not}. Our study also reports water footprint results adjusted for regional water stress. Karimi et al. examined the trade-offs between water usage and energy consumption in data centers as influenced by their cooling system configurations~\cite{karimi2022water}. In our study, we explore the trade-offs between water footprint and carbon footprint based on the geographic location of the GitHub Actions runners.

\emph{Carbon and Water footprints in other fields.} Research in various fields has explored the carbon footprint of computation within specific domains.
For example, in machine learning, Faiz et al. and Luccioni et al. estimated the carbon footprint of training large language models~\cite{faiz2023llmcarbon,luccioni2023estimating}.
Grealey et al. estimated the carbon footprint of bioinformatic tools and commonly run analyses~\cite{grealey2022carbon}. Zuccon et al. and Herrera et al. estimated the water footprint associated with AI infrastructure~\cite{zuccon2023beyond, herrera2025sustainable}.

\section{Conclusion}
In this paper, we estimate the carbon and water footprints of the GitHub Actions ecosystem.

We use Github-provided data for
the execution time of workflow runs, along with estimates for the average CPU, memory, and network usage. These metrics are derived by reexecuting real-world workflows.

Since GitHub does not provide specific information about the regions where the runners are deployed, we account for estimations across different regions. In 2024, our estimates for the carbon footprint of the GitHub Actions ecosystem range from \emissionsOptimisticScenario MTCO2e, if runners are deployed in \textit{Norway West}, to \emissionsPessimisticScenario MTCO2e, if runners are deployed in \textit{India}. In our most likely scenario where runners are deployed in the US, the carbon footprint is projected to be \emissionsLikelyScenario MTCO2e. This is roughly equivalent to the emissions produced by frying \frangos kg of chicken in an air fryer.

Regarding the water footprint, our estimates range from \waterOptimisticScenario kiloliters, if runners are deployed in \textit{Ireland}, to \waterPessimisticScenario kiloliters, if are runners deployed in \textit{New Zealand}. For our most likely scenario, the water footprint is estimated to be \waterLikelyScenario kiloliters, which is equivalent to \glasses glasses of water.

Finally, we suggest strategies to reduce the environmental impact of the GitHub Actions ecosystem, such as deploying runners in regions with a good trade-off between water consumption and carbon emissions and reducing the size of repositories.

Future work should focus on analyzing the evolution of the carbon footprint of the GitHub Actions ecosystem over time to assess whether the problem is worsening and to estimate its future values. Since GitHub Actions run data is retained for a maximum of 400 days~\cite{GithubRetention-2025}, any longitudinal analysis must be performed annually, which is why we do not include such an analysis in this study. In addition, more research is needed to identify and evaluate new strategies to reduce the carbon footprint of CI/CD runs.

\section{Data Availability}
We provide the scripts and dataset used in this paper here: \url{https://doi.org/10.5281/zenodo.16619699}.

\section*{Acknowledgments}

This work was supported by Fundação para a Ciência e a Tecnologia (FCT): N. Saavedra by grant BD/04736/2023 (https://doi.org/10.54499/2023.04736.BD);
N. Saavedra and J.\,F. Ferreira by
projects UID/50021/2025 and UID/PRR/50021/2025 and the `InfraGov' project, with ref. n. 2024.07411.IACDC (DOI: 10.54499/2024.07411.IACDC), funded by the `Plano de Recuperação e Resiliência (PRR)' under the investment `RE-C05-i08 - Ciência Mais Digital', measure `RE-C05-i08.M04' (in accordance with the FCT Notice No. 04/C05 i08/2024), framed within the financing agreement signed between the `Estrutura de Missão Recuperar Portugal (EMRP)' and the FCT as an intermediary beneficiary.
A. Mendes was funded by national funds through FCT – Fundação para a Ciência e a Tecnologia, I.P., under the support UID/50014/2023 (https://doi.org/10.54499/UID/50014/2023).
Icons in Figures~\ref{fig:emissions_comparison}, \ref{fig:water_comparison}, and~\ref{fig:methodology} were made by Freepic, Flaticon, Pixel perfect and max.Icons from www.flaticon.com.

\bibliographystyle{IEEEtran}
\bibliography{references}

\end{document}